\documentclass[12pt]{article}
\usepackage{cite}
\usepackage{bm}
\usepackage{dcolumn}  
\newcommand{\be}{\begin{equation}}
\newcommand{\ee}{\end{equation}}
\newcommand{\bea}{\begin{eqnarray}}
\newcommand{\eea}{\end{eqnarray}}
\newcommand{\beqar}{\begin{eqnarray*}}
\newcommand{\eeqar}{\end{eqnarray*}}

\begin{document}
\baselineskip 18pt%
\begin{titlepage}
\vspace*{1mm}%
\hfill%
\vspace*{15mm}%
\hfill
\vbox{
    \halign{#\hfil         \cr
%           hep-th/yymmnnn\cr
         IPM/P-2009/026  \cr
          } % end of \halign
      }  % end of \vbox
\vspace*{10mm}
\centerline{{\Large {\bf Extremal rotating solutions in Ho\v{r}ava Gravity }}}
\centerline{{\Large{\bf    }}}
\vspace*{5mm}
\begin{center}
{ Ahmad Ghodsi$^a$ and Ehsan Hatefi$^{a,b}$}\\
\vspace*{0.2cm}
$^a$ { Department of Physics, Ferdowsi University of Mashhad, \\
P.O. Box 1436, Mashhad, Iran}\\
\vspace*{0.1cm}
%and\\
$^b$ { School of Physics, 
 Institute for research in fundamental sciences (IPM), \\
 P.O. Box 19395-5531, Tehran, Iran. 
}\\\vspace*{1.5cm}
\end{center}

\begin{abstract} 
Recently a new four-dimensional non relativistic renormalizable theory of gravity was proposed by Horava. In this paper we have found different near horizon geometries in Horava gravity. We find the rotating solutions in a special range of parameters of the theory. This range allows us to include up to six derivative terms in the equations of motion. We have found our solutions as perturbed solutions over spherical symmetric background with small rotating parameter.
\end{abstract} 

\end{titlepage}

%%%%%%%%%%%%%%%%%%%%%%%%%%%%%%%%%%%%%%%%%%%%%%%%%%%%%%%%%%%%%%%%%%%%%%%%%
\section{Introduction to Ho\v{r}ava Gravity}
Recently a new four-dimensional non relativistic renormalizable theory of gravity  was proposed by
Ho\v{r}ava \cite{Horava:2009uw}. It is believed that this theory is a  UV completion for the Einstein theory of gravitation. A great deal of efforts have been done to understand this theory \cite{Lu:2009em},\cite{Horava:2008ih}. In this section we try to review the Ho\v{r}ava gravity and its equations of motion. In next sections we will find solutions to the equations of motion. Our solutions include the extremal spherical and rotating black holes of Ho\v{r}ava gravity.

We start from the four-dimensional metric written in the ADM formalism,
\bea
ds_4^2&=&- N^2  dt^2 + g_{ij} (dx^i - N^i dt) (dx^j - N^j dt)\,.
\label{2}\eea
The Einstein-Hilbert action is given by
\bea
S_{EH} &=& \frac{1}{16\pi G} \int d^4x \sqrt{g} N (K_{ij} K^{ij} - K^2 + R - 2\Lambda)\,,
\eea
where $G$ is the four dimensional Newton's constant and $K_{ij}$ is the second fundamental form and is defined by
\bea
K_{ij} = \frac{1}{2N} (\partial_t g_{ij} - \nabla_i N_j - \nabla_j N_i)\,.
\eea
The action proposed by Ho\v{r}ava describes a non-relativistic renormalizable theory of gravitation and is given by the following action \cite{Horava:2009uw}
\bea
S&=&\int dtd^3x\sqrt{g}N\left\{\frac{2}{\kappa^2}(K_{ij}K^{ij}
-\lambda K^2)+\frac{\kappa^2\mu^2(\Lambda_W R
-3\Lambda_W^2)}{8(1-3\lambda)}+\frac{\kappa^2\mu^2
(1-4\lambda)}{32(1-3\lambda)}R^2\right.\,\nonumber\\
&&\left.\quad\quad\quad\quad\quad\quad-\frac{\kappa^2\mu^2}{8}R_{ij}R^{ij}
+\frac{\kappa^2\mu}{2w^2}\epsilon^{ijk}R_{i\ell}\nabla_jR_k^\ell
-\frac{\kappa^2}{2w^4}C_{ij}C^{ij}\right\}\,,\label{act}
\eea
where $\lambda\,,\kappa\,,\mu\,,w$ and $\Lambda_W$ are constant
parameters, and $C_{ij}$ is the Cotton tensor, defined by
\bea
C^{ij}=\epsilon^{ik\ell}\nabla_k\left(R^j{}_\ell
-\frac14R\delta_\ell^j\right)=\epsilon^{ik\ell}\nabla_k R^j{}_\ell
-\frac14\epsilon^{ikj}\partial_kR\,.
\eea
Using the relation 
\bea
\epsilon^{ijk}R_{i\ell}\nabla_jR^\ell{}_k=R_{i\ell}\left[C^{i\ell}
+\frac14\epsilon^{ij\ell}\partial_j R\right]=R_{i\ell}C^{i\ell}\,,\eea
one can rewrite the action (\ref{act}) as
\bea
S&=&\int dtd^3x\, ({\cal L}_0 + {\cal L}_1)\,,\nonumber\\ {\cal L}_0
&=& \sqrt{g}N\left\{\frac{2}{\kappa^2}(K_{ij}K^{ij} -\lambda
K^2)+\frac{\kappa^2\mu^2(\Lambda_W R
  -3\Lambda_W^2)}{8(1-3\lambda)}\right\}\,,\nonumber\\ {\cal L}_1&=&
\sqrt{g}N\left\{\frac{\kappa^2\mu^2 (1-4\lambda)}{32(1-3\lambda)}R^2
-\frac{\kappa^2}{2w^4} \left(C_{ij} -\frac{\mu w^2}{2}R_{ij}\right)
\left(C^{ij} -\frac{\mu w^2}{2}R^{ij}\right)\right\}\,.\label{action}
\eea% 
By Comparing ${\cal L}_0$ with the general theory of relativity in the ADM formalism, one can read the speed of light \footnote{In this paper we consider $c=1$.}, Newton's constant and the cosmological constant as
\bea
c=\frac{\kappa^2\mu}{4} \sqrt{\frac{\Lambda_W}{1-3\lambda}}\,,\qquad
G=\frac{\kappa^2}{32\pi\,c}\,,\qquad
\Lambda=\frac32 \Lambda_W\,.
\eea
Additionally, demanding that ${\cal L}_0$ gives the usual four dimensional Einstein-Hilbert Lagrangian (general covariance), one finds that $\lambda=1$.

The equations of motion for the action (\ref{action}) are given in \cite{Lu:2009em}. By variation of the Lagrangian with respect to $N$ we find
\be
\frac{2}{\kappa^2}(K_{ij}K^{ij} -\lambda K^2) -
\frac{\kappa^2\mu^2(\Lambda_W R
-3\Lambda_W^2)}{8(1-3\lambda)} -\frac{\kappa^2\mu^2
(1-4\lambda)}{32(1-3\lambda)}R^2 + \frac{\kappa^2}{2w^4}
Z_{ij} Z^{ij}=0\,,\label{EOMN}
\ee
where
\be
Z_{ij}\equiv C_{ij} - \frac{\mu w^2}{2} R_{ij}\,.
\ee
The variation of $N^i$ gives the following equation of motion
\be
\nabla_k(K^{k\ell}-\lambda\,Kg^{k\ell})=0\,.\label{EOMNi}
\ee
and finally, the equations of motion due to the variations with respect to $g^{ij}$ are given by
\be
\frac{2}{\kappa^2}E_{ij}^{(1)}-\frac{2\lambda}{\kappa^2}E_{ij}^{(2)}
+\frac{\kappa^2\mu^2\Lambda_W}{8(1-3\lambda)}E_{ij}^{(3)}
+\frac{\kappa^2\mu^2(1-4\lambda)}{32(1-3\lambda)}E_{ij}^{(4)}
-\frac{\mu\kappa^2}{4w^2}E_{ij}^{(5)}
-\frac{\kappa^2}{2w^4}E_{ij}^{(6)}=0\,,\label{EOMEij}
\ee
where
\bea
E_{ij}^{(1)}&=&
N_i \nabla_k K^k{}_j + N_j\nabla_k K^k{}_i -K^k{}_i \nabla_j N_k-
   K^k{}_j\nabla_i N_k - N^k\nabla_k K_{ij}\nonumber\\&& 
   - 2N K_{ik} K_j{}^k
  -\frac{1}{2} N K^{k\ell} K_{k\ell}\, g_{ij} + N K K_{ij} + \dot K_{ij}
\,,\nonumber\\
E_{ij}^{(2)}&=& \frac{1}{2} NK^2 g_{ij}+ N_i \partial_j K+
N_j \partial_i K- N^k (\partial_k K)g_{ij}+  \dot K\, g_{ij}\,\,,\nonumber\\
E_{ij}^{(3)}&=&N(R_{ij}-
\frac{1}{2}Rg_{ij}+\frac{3}{2}\Lambda_Wg_{ij})-(
\nabla_i\nabla_j-g_{ij}\nabla_k\nabla^k)N\,,\nonumber\\
E_{ij}^{(4)}&=&NR(2R_{ij}-\frac{1}{2}Rg_{ij})-
2 \big(\nabla_i\nabla_j
-g_{ij}\nabla_k\nabla^k\big)(NR)\,\,,\nonumber\\
E_{ij}^{(5)}&=&\nabla_k\big[\nabla_j(N Z^k_{~~i})
+\nabla_i(N Z^k_{~~j})\big]  -\nabla_k\nabla^k(NZ_{ij})
-\nabla_k\nabla_\ell(NZ^{k\ell})g_{ij}\,\,,\nonumber\\
E_{ij}^{(6)}&=&-\frac{1}{2}NZ_{k\ell}Z^{k\ell}g_{ij}+
2NZ_{ik}Z_j^{~k}-N(Z_{ik}C_j^{~k}+Z_{jk}C_i^{~k})
+NZ_{k\ell}C^{k\ell}g_{ij}\nonumber\\&&
-\frac{1}{2}\nabla_k\big[N\epsilon^{mk\ell}
(Z_{mi}R_{j\ell}+Z_{mj}R_{i\ell})\big]
+\frac{1}{2}R^n{}_\ell\, \nabla_n\big[N\epsilon^{mk\ell}(Z_{mi}g_{kj}
+Z_{mj}g_{ki})\big]\nonumber\\&&
-\frac{1}{2}\nabla_n\big[NZ_m^{~n}\epsilon^{mk\ell}
(g_{ki}R_{j\ell}+g_{kj}R_{i\ell})\big]
-\frac{1}{2}\nabla_n\nabla^n\nabla_k\big[N\epsilon^{mk\ell}
(Z_{mi}g_{j\ell}+Z_{mj}g_{i\ell})\big]\nonumber\\&&
+\frac{1}{2}\nabla_n\big[\nabla_i\nabla_k(NZ_m^{~n}\epsilon^{mk\ell})
g_{j\ell}+\nabla_j\nabla_k(NZ_m^{~n}\epsilon^{mk\ell})
g_{i\ell}\big]\nonumber\\&&
+\frac{1}{2}\nabla_\ell\big[\nabla_i\nabla_k(NZ_{mj}
\epsilon^{mk\ell})+\nabla_j\nabla_k(NZ_{mi}
\epsilon^{mk\ell})\big]-\nabla_n\nabla_\ell\nabla_k
(NZ_m^{~n}\epsilon^{mk\ell})g_{ij}\,.
\eea
We will use these equations of motion in order to find new solutions in Ho\v{r}ava gravity. 
%%%%%%%%%%%%%%%%%%%%%%%%%%%%%%%%%%%%%%%%%%%%%%%%%%%%%%%%%%%%%%%%%%%%%%%%%%%%%%%%%%%%%%%%%%%%%%%%%%%%
\section{Extremal spherical solutions}

Before we start to studying the rotating solutions we are interested to find the near horizon behavior of the known static, spherical symmetric solutions. These solutions will be the special case of our solutions when we set the rotation parameter to zero. The static, spherical symmetric solutions has been found in \cite{Lu:2009em}. Similar to \cite{Lu:2009em} we start from the following ansatz 
\be
ds^2 = - N(r)^2\,dt^2 + \frac{dr^2}{f(r)} + r^2 (d\theta^2
+\sin^2\theta d\phi^2)\,.
\ee
If we consider only the Lagrangian ${\cal L}_0$, then we will find the (A)dS Schwarzschild black hole as the following
\be
N^2=f=1 - \frac{\Lambda_Wr^2}{2}  - \frac{M}{r}\,.
\ee
The above solution has an extremal limit with critical mass $M_{0}=\frac{2\sqrt{6}}{9\sqrt{\Lambda_W}}$ and a horizon located at $r_0=\frac{\sqrt{6}}{3\sqrt{\Lambda_W}}$. The near horizon of such a geometry is given by 
%\be
%N^2=f=1 - \frac{\Lambda_Wr^2}{2}  - \frac{M}{r}=(r-r_0)^2(-\frac{\Lambda_W r_o}{r}-\frac{\Lambda_W }{2})
%\ee
%Now using change of variable as $r=r_0+\frac{\lambda}{y}$  and taking the limit $\lambda$ goes to zero we can find $g_{tt}$ and $g_{rr}$
%as the following
\be
ds^2=\frac{3\Lambda_W}{2}\frac{dt^2}{r^2}-\frac{2}{3\Lambda_W}\frac{dr^2}{r^2}+\frac{2}{3\Lambda_W}(d\theta^2+\sin^2\theta d\phi^2)\label{sol1}
\ee
If we consider the total Lagrangian ${\cal L}_0+{\cal L}_1$, then will we find  the following solution in the special case of $\lambda=1$
\be N^2=f=1 - \Lambda_W r^2 - M\, (-\Lambda_W)^\frac14 r^\frac12\,.
\ee
The extremal solution corresponding for this case can be found when we consider the value of $M$ to be $M_0=\frac{4}{3^\frac34}$. Also the horizon is located at $r_0=\frac{1}{\sqrt{-3\Lambda_W}}$.
The near horizon geometry of such a metric is given by
\be
ds^2=-\frac{2(-3\Lambda_W)^{\frac12}}{r^2}dt^2-\frac{2}{3\Lambda_Wr^2}dr^2-\frac{1}{3\Lambda_W}(d\theta^2+\sin^2\theta d\phi^2)\,.
\ee
One can find the solution for general value of $\lambda$. The solution to the equations of motion is 
\bea
&&f=1-\Lambda_W r^2-M ((-\Lambda_W)^\frac12 r)^m\,,\quad N^2=((-\Lambda_W)^\frac12 r)^k f\,,\cr &&\cr
&& m=\frac{2\lambda-\sqrt{6\lambda-2}}{\lambda-1}\,,\quad k=2(1-2m)\,.
\eea
If we impose the reality condition for $m$ then $\frac13\le\lambda<\infty$ or equivalently $-1\le m\le 2$.
Once again in order to find the extremal solution we need to find the critical value for $M$ and the location of the horizon. These are given as follows
\be
M_0=2 m^{-\frac{m}{2}}(2-m)^{-1+\frac{m}{2}}\,,\quad r_0=\sqrt{\frac{m}{(m-2)\Lambda_W }}\,,
\ee
Considering $r_0$ as a real parameter, we note that for $\Lambda_W<0$, then $0\le m \le 2$ and for $\Lambda_W>0$ we have $-1\le m \le 0$.  By considering these constraints we find the near horizon of the above geometry by using the following change of the variables
\be
r\rightarrow \bigg(r_0^m+\frac{\epsilon}{r}\bigg)^{\frac{1}{m}}\,,\quad t\rightarrow \frac{t}{\epsilon }\,.
\ee
Sending $\epsilon\rightarrow 0$ one finds the near horizon metric as 
\bea
ds^2=-\frac{(-\Lambda_W)^m}{m}(\frac{m}{2-m})^{1-3m}\frac{dt^2}{r^2}+\frac{1}{(m-2)\Lambda_W}\frac{dr^2}{r^2}+\frac{m}{(m-2)\Lambda_W}(d\theta^2+\sin^2\theta d\phi^2)\,.\label{NHEl}
\eea
By a scaling of time one may rewrite the above solution as a product space of $AdS_2\times S^2$ with different radii
\be
ds^2=\frac{1}{(m-2)\Lambda_W}\bigg(\frac{-dt^2+dr^2}{r^2}+m(d\theta^2+\sin^2\theta d\phi^2)\bigg)\,.
\ee
%%%%%%%%%%%%%%%%%%%%%%%%%%%%%%%%%%%%%%%%%%%%%%%%%%%%%%%%%%%%%%%%%%%%%%%%%%%%%%%%%%%%%%%%%%%%%%%%
\section{Extremal rotating solutions}
In this section we are interested to find new solutions of Ho\v{r}ava gravity which include angular momentum. We divide our work into three parts. In first part we consider those terms in the Lagrangian which give rise to the two derivative terms in the equations of motion. In the second part we improve our work and consider up to fourth derivative terms in the equations of motion by sending $w\rightarrow\infty$. Finally we will obtain the solution to the equations of motion by taking into account the Cotton tensor, which in this case we have six derivative terms in equations of motion.
\subsection{Two derivative terms}
If one consider up to second derivative terms in the Lagrangian, one finds the known topological rotating solutions given by \cite{Klemm:1997ea} for equations of motion. In the Einstein gravity its metric reads in Boyer-Lindquist type coordinates as the following
\bea
ds^2 = -{\frac{\Delta_r}{\Xi^2\rho^2}}\left[dt-a\sin^2\theta\ d\phi\right]^2
+{\frac{\rho^2}{\Delta_r}} dr^2+{\frac{\rho^2}{\Delta_\theta}}d\theta^2
+{\frac{\Delta_\theta\sin^2\theta}{\Xi^2\rho^2}}
\left[adt-(r^2+a^2)d\phi\right]^2\,,
\eea
where
\bea 
\rho^2&=&r^2+a^2\cos^2\theta\,,\qquad\Xi=1-{a^2\over l^2}\,,\nonumber\\
\Delta_r&=&(r^2+a^2) (1+{r^2\over l^2})-2 M r \,, \qquad \Delta_\theta=1-{a^2\over l^2}\cos^2\theta \,,
\eea
and $a$ is the rotational parameter. Here we consider $l^2=-\frac{2}{\Lambda_W}$ in our notation. One may also write the above metric in the ADM form \cite{Klemm:1997ea} as
\bea
ds^2 = -{\frac{\rho^2\Delta_r\Delta_\theta}{\Xi^2\Sigma^2}}dt^2+{\frac{\rho^2}{\Delta_r}} dr^2+{\frac{\rho^2}{\Delta_\theta}}d\theta^2
+{\frac{\Sigma^2\sin^2\theta}{\Xi^2\rho^2}}(d\phi-\varpi dt)^2\,,
\eea
where
\be
\Sigma^2=(r^2+a^2)^2\Delta_\theta-a^2\sin^2\theta \Delta_r\,,\quad \varpi=-\frac{a}{\Sigma^2}\left[-(r^2+a^2)\Delta_\theta+\Delta_r\right]\,.
\ee
Writing in this form one can explicitly check that it satisfies the equations of motion (\ref{EOMN}), (\ref{EOMNi}) and (\ref{EOMEij}) at the level of two derivative terms. 

Again we are interested to find the near horizon geometry of this solution. The extremal solution happens at a critical mass $M_0=\frac{r_0}{l^2}(l^2+2r_0^2+a^2)$, where the parameter
\be
{r_0}=\sqrt{\frac16(-a^2-l^2\pm\sqrt{a^4+14 a^2 l^2+l^4})}
\ee
gives the location of the horizon. We must note that in the limit where $a\rightarrow 0$, the solution with "$+$" sign has a naked singularity but the the other one with "$-$" sign gives our previous result in solution (\ref{sol1}).

In order to find the near horizon geometry of the extremal solution we define new dimensionless coordinates as
\bea
t\rightarrow \frac{r_0\Xi(r_0^2+a^2)}{2 a^2 }\frac{t}{\epsilon}\,,\quad r\rightarrow r_0+\frac{\epsilon a}{r}\,, \quad \phi\rightarrow \Xi(\frac{r_0^2}{r_0^2+a^2}\phi+\frac{r_0}{2 a}\frac{t}{\epsilon})\,.
\eea
If we send $\epsilon\rightarrow 0$, while keeping the new coordinates $(t,r,\theta,\phi) $ fixed, then
the result will be the following geometry
\bea
ds^2&=&2 r_0^2\Omega^2(\theta)\bigg(-\frac14 \frac{r_0^2}{l^2} f_0\frac{dt^2}{r^2}+\frac{1}{f_0}\frac{dr^2}{r^2}+\frac{1}{\Delta_\theta}d\theta^2+\Delta_\theta\Lambda^2(\theta)(d\phi+\frac{a}{l\, r}dt)^2\bigg)\,,\label{NHK}
\eea
where we have scaled time as $t\rightarrow \frac{a}{l}\, t$ and
\bea
\Omega^2(\theta)=\frac12(1+\frac{a^2}{r_0^2}\cos^2\theta)\,,\quad \Lambda(\theta)=\frac{\sin\theta}{1+\frac{a^2}{r_0^2}\cos^2\theta}\,,\quad f_0=1+\frac{6r_0^2+a^2}{l^2}\,.
\eea
We can do time scaling more to rewrite the first two parts in the metric as an $AdS_2$ metric.
%%%%%%%%%%%%%%%%%%%%%%%%%%%%%%%%%%%%%%%%%%%%%%%%%%%%%%%%%%%%%
\subsection{Four derivative terms}
After finding the exact solution for equations of motion for second order derivative Lagrangian, we are interested in considering the higher derivative terms. Because the equations of motion contain up to six derivative terms, it is very difficult to find the exact solutions. In order to understand the behavior of rotating solutions in Horava gravity we try to solve the equations of motion for a special case, when
the real parameter $w\rightarrow \infty$. Doing this, is similar to the fact that we ignore the effect of Cotton tensor in our equations of motion. Even in this case it is difficult to solve the equations of motion exactly. Hopefully we can find our solutions as the perturbed solution over the spherical solutions by considering the rotating parameter to be small i.e. $a<<l$. So in fact our solution will describe the near horizon solution of slow rotating extremal Kerr black holes in Horava gravity.

Using the near horizon geometry for topological rotating black holes found in (\ref{NHK}) and by changing of the variable as $x=\cos\theta$ we can make a general ansatz for an extremal rotating black hole as follow
\be
ds^2=-N^2(x) \frac{dt^2}{l^2\,r^2}+ A_{1}(x) \frac{dr^2}{r^2}+A_2(x) \frac{dx^2}{1-x^2}+A_3(x) (1-x^2) (d\phi+\frac{a}{l\,r} dt)^2\,,
\ee 
This ansatz automatically satisfies in equation (\ref{EOMNi}). There are some points about our ansatz. Because there are two possible field redefinitions for $x$ and $t$ one can use them to fix the value of $g_{xx}$ and the scaling of time. The former can be fixed by choosing
\be
A_2(x)=r_0^2 (1+\frac{a^2}{r_0^2} x^2) \frac{1}{1-\frac{a^2}{l^2} x^2}\,,\label{A2}
\ee
and the later is fixed by choosing the coefficient $\frac{a}{l\,r}$ inside the off-diagonal term.
The form of equation (\ref{A2}) will help us to find the value of horizon $r_0$ for our extremal solution. One may chooses other values for fixing the ambiguities in fields, but the results will be equivalent with our solution by a field redefinition on the coordinates $x$ and $t$. 

If we start to solve the equations of motion in terms of the rotation parameter $a$, we will find polynomial solutions with some constants of integrations. Most of these constants can be fixed by applying the following constraints.

1. One must notice that in order to solve the equations of motion we need to impose some regularity conditions at poles located at $\theta=0,\pi$ or equivalently at $x=\pm 1$. We use the similar conditions noted in \cite{Astefanesei:2006dd}. It is important to note that these regularity conditions are imposed at the level of $\Lambda_W\rightarrow 0$,
\bea
&&N(x) A_1^{\frac12}(x)\rightarrow constant\,, \quad x\rightarrow \pm1\,, \nonumber\\
&&\frac{A_2(x)}{A_3(x)}\rightarrow 1\,, \quad x\rightarrow \pm1\,.
\eea
The second constraint can be found equivalently by imposing the value of $g_{\phi\phi}=r_0^2$ at $x=0$.
But if we consider a non zero value for $\Lambda_W$ the ratio of $\frac{A_2(x)}{A_3(x)}$ in second constraint will tend to another value, $1+{\mathcal{O}}(\Lambda_W)$. But we still can use the relation $g_{\phi\phi}(x=0)=r_0^2$.

2. We have a symmetry as $x\leftrightarrow -x$, so that the solutions will be even polynomials of $x$.

By knowing the above facts and for exercise let us come back to the case with two derivatives in equations of motion and impose the above constraints. We find the following solution
\bea
N^2(x)&=&c_1^2+\frac{1}{12}l^2(c_2+  x^2)a^2+{\mathcal{O}}(a^4)\,,\nonumber\\
A_1(x)&=&\frac{l^2}{3}+\frac{1}{36}(1+x^2)\frac{l^4}{c_1^2}a^2+{\mathcal{O}}(a^4)\,,\nonumber\\
A_3(x)&=&-\frac{l^2}{3}+\frac{1}{36}\bigg(12 (x^2-1)+\frac{l^4}{c_1^2}(x^2+1)\bigg) a^2+{\mathcal{O}}(a^4)\,,\nonumber\\
r_0&=&\sqrt{-\frac{l^2}{3}}+\frac{1}{72}\sqrt{-\frac{3}{l^2}}(12-\frac{l^4}{c_1^2})a^2\,.\label{2d}
\eea
This solution must result the near horizon spherical symmetric solutions (\ref{NHEl}) when $a=0$ up to a scaling of time, i.e. except the value of $g_{tt}$, the other parameters of the metric are equal. This comes from the fact that the scaling of time in metric (\ref{NHEl}) and the metric (\ref{NHK}) are not the same.
The other two constants $c_1$ and $c_2$ can not be fixed. In fact we can read them by finding the series expansion of (\ref{NHK}). We find the following values for the remaining constants
\be
c_1=\sqrt{-\frac{l^4}{36}}\,,\quad c_2=-5\,.
\ee

We now consider the Horava gravity with four derivative terms in equations of motion. In this case one may note that we have not a similar solution like the metric in (\ref{NHK}). Imposing the regularity and symmetry conditions gives
the following solution to the equations of motion for a general value of  $\lambda$ 
\bea
N^2&=&c_1^2+\frac{2{l}^{2}{\zeta }^{2}}{3{ \left( 5+22\,{\zeta }^{2} \right)  \left( 6\,{\zeta }^{2}+1 \right) ^{2} \left( 2
\,{\zeta }^{2}+1 \right) ^{2}}}\bigg\{1488\,{\zeta }^{8}+1024{\,{\zeta }^{6}+196\,{\zeta }^{4}-6\,{\zeta }^{2}-3}\cr &&\cr
&-&3\,\left({  432\,{\zeta }^{8}+344\,{\zeta }^{6}+52\,{\zeta }^{4}-16\,{\zeta }^{2}-3  }\right)x^2\bigg\}a^2+{\mathcal{O}}(a^4)\,,\cr &&\cr
A_1(x)&=&\frac14 l^2(1+2\zeta^2)-\frac{\zeta^4 l^4}{c_1^2\,(528\zeta^6+472\zeta^4+124\zeta^2+10)}\bigg\{1+28\zeta^4+12\zeta^2\cr &&\cr
&+&\left(2+6\zeta^2+4\zeta^4\right)x^2\bigg\}a^2+{\mathcal{O}}(a^4)\,,\cr &&\cr
A_3(x)&=&\zeta^2 l^2+\frac{1}{4(264\zeta^6+236\zeta^4+62\zeta^2+5)c_1^2}\bigg\{\zeta^4(64\zeta^4+12\zeta^2-1)l^4\cr &&\cr
&+&(-528\zeta^8-1000\zeta^6-596\zeta^4-134\zeta^2-10)c_1^2\cr &&\cr
&+&\bigg(\zeta^4(16\zeta^4-4\zeta^2-3)l^4+(528\zeta^8+1000\zeta^6+596\zeta^4+134\zeta^2+10)c_1^2\bigg)x^2\bigg\}\cr &&\cr
&+&{\mathcal{O}}(a^4)\,,\label{4dg}
\eea
with
\be
r_0=\zeta l-\frac{(\zeta^4-12\zeta^6-64\zeta^8)l^4+(10+134\zeta^2+596\zeta^4+1000\zeta^6+528\zeta^8)c_1^2}{8l\zeta c_1^2(264\zeta^6+236\zeta^4+62\zeta^2+5)}a^2+{\mathcal{O}}(a^4)\,,
\ee
where $\zeta=\sqrt{\frac16(-1+\sqrt{-2+6\lambda})}$. As we see all the constants have been fixed except $c_1$. We can't fix it because our solution has not the same time scaling as (\ref{NHEl}), but one can easily check that when $a=0$ we can find $g_{rr}$, $g_{\theta\theta}$ and $g_{\phi\phi}$ as in (\ref{NHEl}). 

In the special case of $\lambda=1$ and using the value $\zeta=\frac{1}{\sqrt{6}}$, the above solution reduces to 
\bea
N^2(x)&=&c_1^2+\frac{1}{2496}(33+31 x^2)l^2a^2+{\mathcal{O}}(a^4)\,,\nonumber\\
A_1(x)&=&\frac{l^2}{3}-\frac{1}{7488}(17+14x^2)\frac{l^4}{c_1^2}a^2+{\mathcal{O}}(a^4)\,,\nonumber\\
A_3(x)&=&\frac{l^2}{6}+\frac{1}{29952}\bigg(17472(x^2-1)-(29x^2-25)\frac{l^4}{c_1^2}\bigg)a^2+{\mathcal{O}}(a^4)\,,\nonumber\\
r_0&=&\frac{l}{\sqrt{6}}-\frac{1}{59904}(17472-25 \frac{l^4}{c_1^2})\frac{\sqrt{6}}{l}a^2+{\mathcal{O}}(a^4)\,.\label{4d}
\eea
%%%%%%%%%%%%%%%%%%%%%%%%%%%%%%%%%%%%%%%%%%
\subsection{Six derivative terms}
We can include the effect of Cotton tensor, so we perform our calculations at finite value for $w$. To find the solution perturbatively in terms of $a$ we consider the following ansatz for metric components
\bea
N(x)&=& n+(p+q x^2) a^2\,\quad A_1(x)= s +(u +v  x^2) a^2\,\nonumber\\
A_3(x)&=& f +(g +h  x^2) a^2\,\quad r_0=x_0 + x_1 a^2\,.
\eea
Putting them into the equations of motion we will find the a set of algebraic equations presented in Appendix A. There are two more equations which fix the values of $s$ and $x_0$, our computation shows these are independent of $w$ and give the previous result in (\ref{4dg}). The equations of motion in appendix A is very complicated to solve in terms of general value of $\lambda$ parameter like what we found in (\ref{4dg}) so we restrict ourselves to the specific value of $\lambda=1$. By this we can compare our solutions in the Horava gravity with those in the Einstein gravity. The solution to the equations of motion will be
\bea
N^2(x)&=&n^2+\frac{1}{2496}{\frac { 5\left(33+31\,{x}^{2}
 \right) {\omega}^{4}{l}^{4}+162 \left( 22+5\,{x}^{2} \right) {k}^{4}
  }{5\,{\omega}^{4}{l}^{4}+108\,{k}^{4}}}l^2{a}^{2}\,,\cr &&\cr
A_1(x)&=&\frac{{l}^{2}}{3}-\frac{1}{7488}{\frac {5\left( 17+14\,{x}^{2} \right) {
\omega}^{4}{l}^{4}+270 \left( 1+2\,{x}^{2} \right) {k}^{4} }
{5\,{\omega}^{4}{l}^{4}+108\,{k}^{4}}} \frac{l^4\,a^{2}}{\,n^2}\,,\cr &&\cr
A_3(x)&=&\frac{5}{78}{\frac {13\,{\omega}^{4}{l}^{4}+216\,{k}^{4}}{5\,{\omega}^{4}{l}^{4}+108\,{k}^{4}}}l^2+\frac{1}{( 5\,{\omega}^{4}{l}^{4}+108\,{k}^{4} ) ^{2}}\, \bigg[ \frac{25}{29952}\,( 25-29\,{x}^{2} ) {\omega}^{8}{l}^{12}\cr &&\cr
&+&\frac{175}{12} \bigg(  ( -1+{x}^{2} ) {{ n}}^{2}{\omega}^{4}+ ( \frac{45}{728}-\frac{639}{18928}\,{x}^{2} ) {k}^{4} \bigg) {\omega}^{4}{l}^{8}\cr &&\cr
&-&\frac{2025}{1352}\,
 \bigg( {\frac {16744}{45}}\,{{n}}^{2} ( {\frac {26}{23}}-{x}^{2} ) {\omega}^{4}- ( \frac{13}{2}-{x}^{2} ) {k}^{4}
 \bigg) {k}^{4}{l}^{4}
-\frac{68040}{13}\,{{n}}^{2}{k}^{8} ( {
\frac {13}{10}}-{x}^{2} )  \bigg] \frac{a^{2}}{n^{2}}
\,,\cr &&\cr
 r_0&=&\frac{l}{\sqrt{6}}-\frac{1}{59904}(17472-\frac{25{l}^{4}}{n^2})\frac{\sqrt{6}}{l}a^2\,.\label{6d}
\eea
As we see this result gives exactly the four derivatives result in (\ref{4d}) with $n=c_1$ when one sends $w\rightarrow\infty$.
%%%%%%%%%%%%%%%%%%%%%%%%%%%%%%%%%%%%%%%%%%
\section{Discussion}
In this section we try to compare the result of the Einstein gravity for two derivative solution in equation (\ref{NHK}) or (\ref{2d}), with the four derivative (\ref{4d})  and the six derivative (\ref{6d}) solutions of the Horava gravity. We also argue about the properties of our solutions. 

{\bf{1.}}  Limiting ourselves in $\lambda=1$ one can precisely see that it is not possible to recover the result in (\ref{2d}) from (\ref{4d}) or (\ref{6d}). This was first observed in \cite{Lu:2009em} and our results reconfirm this behavior.

{\bf{2.}} One can also compare the location of the horizon $r_0$. As we see this location for solution (\ref{2d}) differs from  (\ref{4d}). Additionally our six derivatives solution (\ref{6d}) shows that the location of horizon is independent of $w$.

{\bf{3.}}
One important parameter for rotating solutions is the angular velocity. If we consider the rotation relative to the stationary frame at infinity then the angular velocity is given by
\be
\Omega_{H}=\frac{a}{l\,r_0}=\frac{a}{\zeta l^2}+{\mathcal{O}}(a^3)\,.
\ee 
This value is given by $\frac{d\phi}{dt}$ along the time-like trajectories of co-rotating observers with fixed values for $r$ and $\theta$. The time $t$ is proportional to the proper time as $t=\frac{l\,r}{N(x)}\tau$.

There is an interesting property for angular velocity for a fixed value of the rotating parameter ``$a$". For $\Lambda_W>0$, the angular velocity increases monotonically in terms of $\lambda$. On the other hand for $\Lambda_W<0$ the angular velocity decreases monotonically.

There is also another interesting point when one considers the (\ref{6d}) solution. As one sees the value of $r_0$ is independent of $w$ and this shows that in this region of parameter space (slow rotation) $w$ is not relevant in the value of angular velocity. 

{\bf{4.}} To recover solution in (\ref{4d}) from (\ref{6d}) we must send $w\rightarrow \infty$ and to recover the non-rotating solutions we must send $a\rightarrow 0$. But the solution in (\ref{6d}) suggests that if we first send $a\rightarrow 0$ then we must also send simultaneously $w\rightarrow \infty$ to recover the stationary solution.

{\bf{5.}} Solution (\ref{6d}) suggests that there is another $w$-independent solution in the region of parameter space when $\kappa\rightarrow \infty$. This solution has the same location of horizon and angular velocity as (\ref{4d}) but one cannot recover the stationary metric  when $a\rightarrow 0$. This also suggests that the region of large $\kappa$ and nonzero value of $a$ is disconnected from the region of finite $\kappa$ and zero value of $a$.
\section*{Acknowledgment}
A. G would like to thank Hong Lu for discussion and E. H would like to thank M. R. Garousi. This work was supported by the grant (P/446, 88/7/21) from Ferdowsi University of Mashhad.
%%%%%%%%%%%%%%%%%%%%%%%%%%%%%%%%%%%%%%%%%%%%%%%%%%%%%%%%%%%%%%%%%%%%%%%%%%
\section*{Appendix A}
\bea
f ^2x_0^5+4\bigg((hs +f v )x_0^3+2f s x_1x_0^2+(\frac12\lambda f v- (\lambda-\frac12)hs  )l^2x_0-2 (\lambda-\frac12)x_1s fl^2\bigg)n^2=0\,,
\eea
\bea
& &f ^2l^2x_0^6+\bigg(-8s f x_0^4+\bigg((8\lambda-4)s f +\bigg(6v f \lambda+(-16h \lambda+8h )s \bigg)x_0^2\bigg)l^4\nonumber\\
&+&8\bigg((\lambda-\frac32)s f +(2s h +\frac32v f )x_0^2\bigg)l^2x_0^2\bigg)n^2=0\,,
\eea
\bea
&-&\frac18\bigg(-12snp-12un^2+fl^2\bigg)fw^4x_0^8+\frac12\bigg((p-2q)sf+(-sh+uf)n\bigg)l^2nw^4x_0^6\nonumber\\
&-&sfl^2n^2x_1w^4x_0^5-\frac14\bigg((-\frac12p+(p+2q)\lambda)sfl^2+\bigg(-4sf\lambda+(6(\lambda+\frac16)sh+(-\frac12u\nonumber\\
&+&5v+(u-9v)\lambda)f)l^2\bigg)n\bigg)l^2nw^4x_0^4+(\lambda-\frac12)sfl^4n^2x_1w
^4x_0^3-\bigg(-sfl^4\lambda w^4-sf\kappa^4\nonumber\\
&+&2(sh-\frac12vf)\kappa^4l^2\bigg)n^2x_0^2+sf\kappa^4l^2n^2=0\,,
\eea
\bea
&-&\frac{1}{24}fw^4\bigg(12vn^2+12snq+fl^2\bigg)x_0^8-\frac16n\bigg(\bigg((-2s+l^2v)f+4l^2sh\bigg)
n+7qsl^2f\bigg)w^4x_0^6\nonumber\\
&+&\frac{1}{12}n\bigg(\bigg(\bigg(28v(-\frac{31}{56}+\lambda)l^2+(6+8\lambda)s\bigg)f-(4+16\lambda)shl^2\bigg)n-5sl^2(\frac{1}{10}+\lambda)qf\bigg)w^4l^2x_0^4\nonumber\\
&-&2n^2\bigg(-(\frac{1}{12}+\frac13\lambda)w^4sfl^4+\kappa^4\bigg((-\frac12l^2v-\frac12s)f+l^2sh\bigg)\bigg)x_0^2+sf\kappa^4l^2n^2=0\,,
\eea
\bea
&+&\bigg(2l^2un^2-2sp(l^2-3s)n+\frac12l^2sf\bigg)w^4fx_0^8-4sn^2fw^4x_1(l^2-3s)x_0^7-\bigg(\bigg(-h\lambda s\nonumber\\
&-&f((u-2v)\lambda+v)\bigg)n+\lambda fsp\bigg)w^4nl^4x_0^6+\bigg(8\bigg(-(\frac14h(\lambda-\frac12)s-\frac18vf\lambda)w^4sl^4\nonumber\\
&+&(sh-\frac12vf)\kappa^4l^2-\frac12sf\kappa^4\bigg)n+pw^4l^4s^2f(\lambda-\frac12)\bigg)nx_0^4-(2\lambda-1)s^2x_1w^4n^2fl^4x_0^3\nonumber\\
&-&2s\bigg((sh-\frac12f(-4+v))l^2-\frac12sf\bigg)\kappa^4n^2x_0^2+l^2n^2s^2f\kappa^4=0\,,
\eea
\bea
&-&\frac16w^4\bigg((-4vn^2+4snq+sf)l^4+(-12ns^2q+4sn^2)l^2-12n^2s^2\bigg)fx_0^8\nonumber\\
&-&\frac13w^4n\bigg(\bigg((4h\lambda s-(4\lambda-2)vf)n+\lambda fsq\bigg)l^4-4s\bigg((-\frac12+\frac14\lambda+\frac12v)n+sq\bigg)fl^2\nonumber\\
&-&6ns^2f\bigg)l^2x_0^6-\frac83n\bigg(\frac38w^4s\bigg((\frac83h(\lambda-\frac12)s-(\frac13(v+1)\lambda+\frac13v)f)n+sfq(\lambda-\frac12)\bigg)l^4\nonumber\\
&+&n\bigg(-(\frac58\lambda-\frac{5}{16})w^4fs^2+\kappa^4hs-\frac12fv\kappa^4\bigg)l^2-\frac12ns\kappa^4f\bigg)l^2x_0^4+2n^2\bigg((\frac56\lambda-\frac{5}{12})w^4sfl^4\nonumber\\
&-&\kappa^4(sh-(\frac23+\frac12v)f)l^2+\frac12sf\kappa^4\bigg)sl^2x_0^2+l^4n^2s^2f\kappa^4=0\,,
\eea
\bea
&-&\bigg(\frac{3}{10}(-\frac43l^2gs+\frac43fl^2u+4gs^2)n^2-\frac25sfp(l^2-3s)n+\frac{3}{10}f^2sl^2\bigg)w^4x_0^8\nonumber\\
&+&\bigg(\frac15\bigg((\lambda(-h+g)l^2+2vf)s-f\lambda(v+u)l^2\bigg)n+\frac15(4sq+l^2\lambda p)
sf\bigg)w^4nl^2x_0^6\nonumber\\
&-&\frac25fw^4sn^2l^4x_1\lambda x_0^5+\frac85n\bigg(\bigg(-(\frac18\lambda-\frac{1}{16})\bigg((14h+g)l^2-8f\bigg)w^4l^2s^2+\bigg(\frac38fv(\lambda+\frac13)w^4l^4\nonumber\\
&+&\kappa^4hl^2-\frac12\kappa^4f\bigg)s-\frac12fl^2v\kappa^4\bigg)n-(\frac12q+\frac18p)w^4(\lambda-\frac12)l^4s^2f\bigg)x_0^4\nonumber\\
&+&\frac25(2\lambda-1)s^2x_1w^4n^2fl^4x_0^3-\bigg(2\bigg(-\frac45fw^4(\lambda-\frac12)l^4+\kappa^4hl^2-\frac12\kappa^4f\bigg)s\nonumber\\
&-&2(-\frac25+\frac12v)\kappa^4fl^2\bigg)sn^2x_0^2+l^2n^2s^2f\kappa^4=0\,,
\eea
\bea
&-&\frac16w^4\bigg((-4hs^2+\frac{4}{3}l^2sh-\frac43fl^2v)n^2+\frac43sfq(l^2-3s)n+f^2sl^2\bigg)x_0^
8\nonumber\\
&-&\bigg(\frac19\bigg((5h\lambda l^2-4f(\frac12\lambda+v))s-fv(-1+\lambda)l^2\bigg)n+\frac19sfq(-8s+l^2\lambda
)\bigg)nw^4l^2x_0^6\nonumber\\
&+&\frac89n\bigg(\bigg(-\frac{23}{8}w^4(l^2h-\frac{12}{23}f)(\lambda-\frac12)l^2s^2+\bigg(\frac58f
((\frac25+v)\lambda+\frac25v)w^4l^4+\kappa^4hl^2-\frac12\kappa^4f\bigg)s\nonumber\\
&-&\frac12fl^2v\kappa^4\bigg)n-\frac78l^4(\lambda-\frac12)w^4s^2qf\bigg)x_0^4-2sn^2\bigg((-\frac23fw^4(\lambda-\frac12)l^4+\kappa^4hl^2-\frac12\kappa^4f)s\nonumber\\
&-&\frac12f(v-\frac49)\kappa^4l^2\bigg)x_0^2+l^2n^2s^2f\kappa^4=0\,,
\eea
%%%%%%%%%%%%%%%%%%%%%%%%%%%%%%%%%%%%%%%%%%%%%%%%%%%%%%%%%%%%%%%%%%%%%%%%% 

\end{document}